 \documentclass[final,5p,times,twocolumn,authoryear,numbers, square, sort&compress]{elsarticle}

\usepackage{epsfig}

\usepackage{amssymb}
\usepackage{lipsum}
\usepackage{amsmath}
\usepackage{amsthm}
\usepackage{graphicx,framed} % Include figure files
\usepackage{dcolumn} % Align table columns on decimal point
\usepackage{bm} % bold math
\usepackage{braket}
\usepackage{dsfont}
\usepackage{color}
\usepackage{amsthm}

\usepackage[normalem]{ulem} %to strike the words
\usepackage{qcircuit}       % professional-quality tables
\usepackage[hidelinks]{hyperref}
\usepackage{amsthm}

\usepackage{comment}
\usepackage{float}
\setcitestyle{numbers}

\makeatletter
\def\ps@pprintTitle{%
     \let\@oddhead\@empty
     \let\@evenhead\@empty
     \def\@oddfoot{\reset@font\hfil}%
     \let\@evenfoot\@oddfoot
}
\makeatother

\usepackage{flushend}

\journal{Annals of Physics}

\begin{document}

\begin{frontmatter}



\title{Time-dependent Aharonov-Bohm type topological effects on dipoles}


\author{H. O. Cildiroglu}
\affiliation{organization={Boston University },%Department and Organization
            addressline={Physics Department}, 
            city={Boston},
            postcode={02100}, 
            state={MA},
            country={USA}}
\affiliation{organization={Ankara University },%Department and Organization
            addressline={Department of Physics Engineering}, 
            postcode={06100}, 
            state={Ankara},
            country={Türkiye}}

\begin{abstract}
Exploring the time-dependent characteristics of AB-type effects holds significant importance in contemporary physics and its practical applications. Here, we delve into the investigation of time-dependent topological effects emerging in AB-type experimental setups. We first analyze the topological effects on magnetic dipoles moving in closed trajectories around the time-varying magnetic field source solenoid, then on electrical dipoles around a time-varying electric field source in 2+1 dimensions without any approximation. Last, we discuss the characteristics of the topological effects by considering the identity and dualities between phases from an integrated perspective.

\end{abstract}




\end{frontmatter}

%\tableofcontents


\section{Introduction}

Quantum mechanical topological effects are elucidated as the introduction of the vector or vector potential-like physical quantities as complex phase factors into the wave functions of particles moving along the closed trajectories around singularities created by electromagnetic field sources without the effects of the classical forces. One of the pioneering and first examples is the Aharonov-Bohm (AB) phase, which demonstrates the significance of electromagnetic potentials in quantum theory ~\cite{Aharonov1959, Chambers1960, Tonomura1986}. This complex phase exerts a measurable impact on the interference pattern of wave functions of electric charges reaching a screen via two distinct trajectories around an infinitely long solenoid as a static magnetic field source. Subsequently, various topological effects have emerged by considering Maxwell's dualities. These include the Aharonov-Casher (AC) effect arises from the motion of magnetic dipoles around a linear electric charge distribution ~\cite{Aharonov1984, Cimmino1989}, the He-McKellar-Wilkens (HMW) effect from the motion of electric dipoles around a linear magnetic charge distribution ~\cite{He1993, Wilkens1994, Lepoutre2013}, and the dual AB effect (DAB) from the movement of magnetic charges around an electric field tube ~\cite{Downing1999, Cildiroglu2018}.

In recent years, studies on AB-type effects have focused on investigating the time dependence of the effect. Singleton et al. discussed two covariant generalizations of the AB effect with time-dependent flux, noting that the AB phase shift is canceled by the phase shift of the external electric field associated with the Lorentz force ~\cite{Singleton2013}, Bright et al. explored the time-dependent (TD) AB effect for non-Abelian gauge fields revealing cancellations between phase shifts from non-Abelian electromagnetic fields ~\cite{Bright2015}, Ababekri et al. examined the non-relativistic behavior of particles with electric dipoles on noncommutative space uncovering quantum phase corrections ~\cite{Ababekri2016}, Singleton et al. developed a covariant expression for the AC phase, investigating its interaction with electromagnetic fields ~\cite{Singleton2016}, Jing et al. re-examined the AB effect in the background of a time-dependent vector potential, highlighting alterations in interference patterns ~\cite{Jing2017}, Ma et al. explored noncommutative corrections to the TD-AB effect by revealing three types of corrections and proposing dimensionless quantities for parameter extraction based on measured phase shifts  ~\cite{Ma2017}, Choudhury et al. discovered a frequency-dependent AB phase shift ~\cite{choudhury2019}, Jing et al. revisited the TD-AB effects in noncommutative space-time, finding no noncommutative corrections to the AB effects for both cases up to the first order of the noncommutative parameter ~\cite{Jing2020}, Wang et al. investigated the TD-HMW effect in noncommutative space by confirming gauge symmetry, and time-dependent AC effect and its corrections due to spatial noncommutativity on noncommutative space ~\cite{Wang2022, Wang2023}, Saldanha proposed an electrodynamic AB scheme challenging the topological nature of the phase ~\cite{Saldanha2023}, and Wakamatsu et al. analyzed the AB effect's interaction energy and its gauge invariance ~\cite{Wakamatsu2024}. 

In this letter, we study the TD-AB type topological effects of time-varying electromagnetic fields on dipoles in 2+1 dimensions. We begin with a TD solenoidal magnetic field oriented in the z direction, and the resulting rotational electric field appearing in the xy plane. To ensure that particles are not subject to classical forces, we use magnetic dipole carrier chargeless particles. Accounting for the relativistic electromagnetic effects arising from the motion of the particles, we give direct calculations for appearance of a vector-potential-like physical quantity which enters as a complex observable phase in the wave functions of the particles. We conclude that the resulting phase is topological. Accordingly, we emphasize that, although there is no direct interaction, the phase involves the time-rate change of the energy term, and the interference pattern can be controlled by means of a changing magnetic field (i.e. flux). Last, we consider the fully dual of this problem and show the topological effects on electric dipole moment carrier chargeless particles moving around a time-varying electric field. 

\section{Time-Dependent AB-type Effect on Magnetic Dipoles}

\begin{figure}
    \begin{minipage}{\linewidth}
        \centering
        \includegraphics[width=\linewidth]{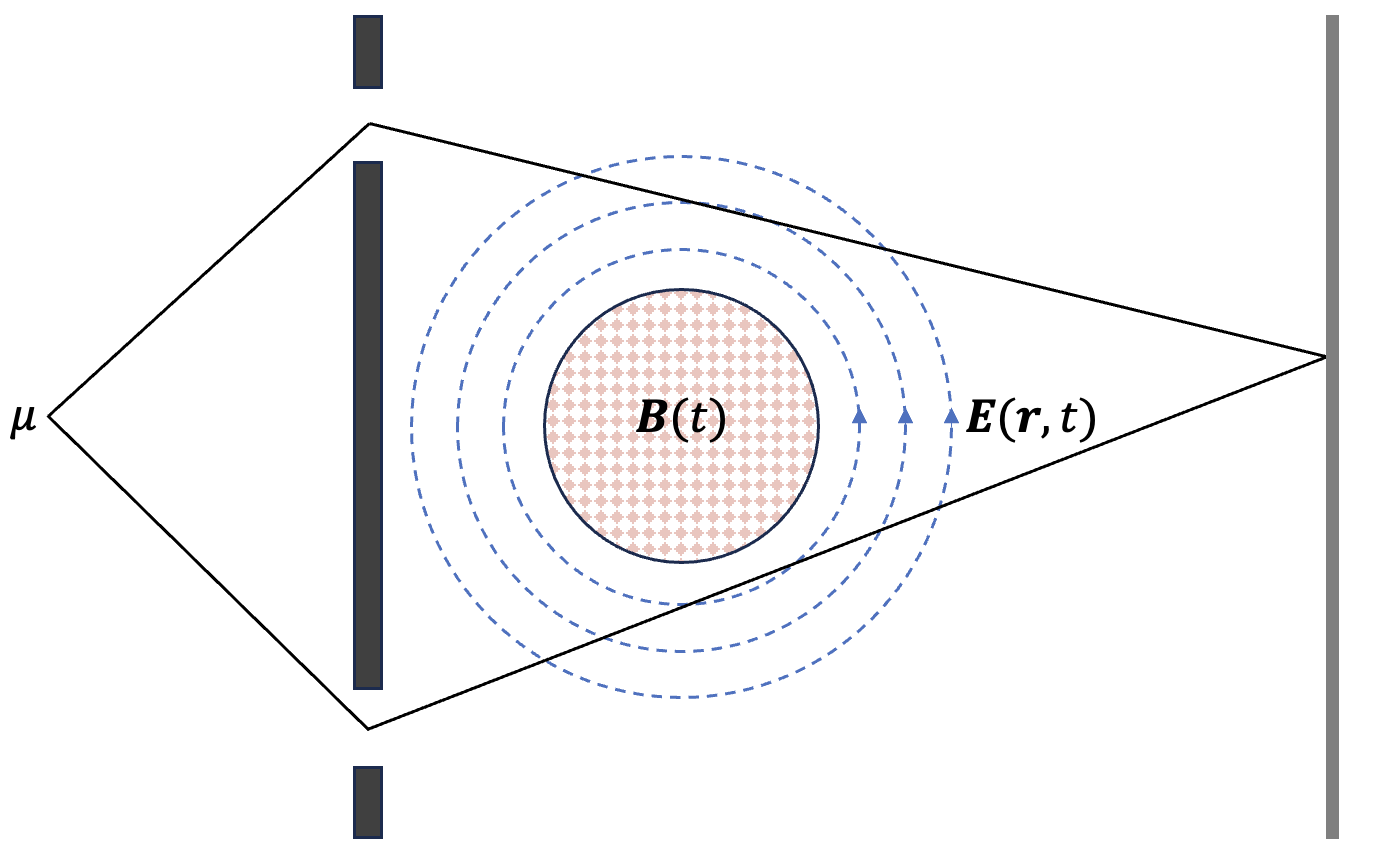}
        \caption{Schematic representation of moving magnetic dipoles around a time-dependent magnetic field source solenoid with the radius $a$ in AB-type setup.}
        \label{fig:AB-magnetic}
    \end{minipage}
\end{figure}

In an AB-type experimental setup, as a time-dependent magnetic field source infinitely long solenoid with the radius of $a$ oriented in the z-direction is placed just behind the double-slits (See Fig.~\ref{fig:AB-magnetic}). In accordance with the nature of AB-type effects and to ensure that the particles are not subject to a classical Lorentz force, we use magnetic dipole moment carrier chargeless particles. In this scenario, the magnetic field within an infinitely long solenoid, one might initially expect the particles to remain unaffected. However, the presence of a time-varying solenoidal field induces a corresponding time-dependent electric field ($\boldsymbol{E}(\boldsymbol{r},t)$) in the xy-plane $({r>a})$. Within this field, a particle moving in an electric field at velocities where first-order effects in terms of $\beta$ (the velocity ratio relative to the speed of light) needs to be considered, is subject to relativistic electromagnetic effects. Vector potential or vector potential-like physical quantities enter the wave functions of the particles as complex phase factors, leading to measurable geometrical and topological effects. Besides, the induced (time-dependent) electric field, in turn, generates a new magnetic field ($\Tilde{\boldsymbol{B}}(\boldsymbol{r},t)$) that interacts with the magnetic dipole ~\cite{Abbott1985}, as anticipated (under appropriate polarization states), and can introduce an additional term to the Hamiltonian, along with a scalar (dynamic) phase contribution to the wave functions of the particles ~\cite{Allman1992}. However, this phase is not a geometrical/topological and can be eliminated since it vanishes in the expectation value expressions. Consequently, it is appropriate to begin with the relativistic Lagrangian that describes the motion of chargeless magnetic dipoles in the presence of electromagnetic fields,

\begin{equation}
    L = \overline{\psi}\left( i\gamma^{\mu}\partial_{\mu} - m + \frac{\mu}{2}\sigma^{\mu\nu}F_{\mu\nu} \right)\psi
    \label{eq:2}
\end{equation}

\noindent and the corresponding equations of motion derived from this Lagrangian:

\begin{equation}
    \left(i\gamma^{\mu}\partial_{\mu} - m + \frac{\mu}{2} \sigma^{\mu\nu}F_{\mu\nu}\right)\psi = 0
    \label{eq:3}
\end{equation}

\noindent Due to the symmetry of the problem, there is invariance under spatial translations along the z-direction, making an examination in 2+1 dimensions sufficient for the emergence of quantum mechanical effects. In 2+1 dimensional spacetime (+,-,-), gamma matrices change from their well-known four-dimensional representations. Therefore, the $\sigma^{\mu\nu}$ term in equation ~\eqref{eq:3} needs to be redefined. With the selection of the free Dirac Hamiltonian $H_{D}$,

\noindent 

\begin{equation}
    H_{D} = \alpha_{x}p_{x} + \alpha_{y}p_{y} + m\beta   
    \label{eq:4}
\end{equation}

\noindent the gamma matrices can be defined in two-dimensional space in two different representations satisfying the Clifford algebra $(s = \pm 1)$ for the up and down spin polarization states:

\begin{equation}
    \begin{matrix}
    \alpha_{x} = \sigma_{x} & & & & \beta = \gamma_{0} = \sigma_{z} & & & & \gamma_{1}=\beta\alpha_{x} = i\sigma_{y}\\
\alpha_{y} = {s\sigma}_{y} & & & &  \{ \gamma^{0}, \gamma^{i} \}=0  & & & &  \gamma_{2}=\beta\alpha_{y} = -is\sigma_{x}
\end{matrix}
\label{eq:5}
\end{equation}

\noindent Hence, by considering the transformation $F_{\mu\nu}=\eta_{\mu\alpha}F^{\alpha\beta}\eta_{\beta\nu}$, the  $\sigma_{\mu\nu}F^{\mu\nu}$ term can be written,

\begin{equation}
    \sigma^{\mu\nu}F_{\mu\nu} = 2\sigma^{0i}F_{0i} + \sigma^{ij}F_{ij} = 2i\left( \sigma_{x}E_{x} + s\sigma_{y}E_{y} \right)+2\sigma_z \Tilde{B}_z 
    \label{eq:6}
\end{equation}

\noindent Here,  $\sigma^{0i} = i\alpha_{i}$, $\sigma^{ij} = \sigma^{k}$. Now, one can multiply ~\eqref{eq:3} by $\gamma^{0}$ from the left after seperating the space and time components of the covariant derivative, it becomes,

\begin{equation}
    \left\lbrack m\beta + \alpha_{x}p_{x} + \alpha_{y}p_{y} + s\mu\left( \alpha_{y}E_{x} - \alpha_{x}E_{y} \right) + \boldsymbol{\mu} \cdot \boldsymbol{\Tilde{B}}\right\rbrack\psi=i\partial_{t}\psi
    \label{eq:7}
\end{equation}

\noindent In this case, it is evident that the expression inside the parentheses on the right-hand side is the total Hamiltonian $(H)$ of the system.

\begin{equation}
    H = \left\lbrack m\beta + \alpha_{x}p_{x} + \alpha_{y}p_{y} + s\mu\left( \alpha_{y}E_{x} - \alpha_{x}E_{y} \right) + \boldsymbol{\mu} \cdot \boldsymbol{\Tilde{B}} \right\rbrack\ 
    \label{eq:8}
\end{equation}

\noindent Thus, by considering ~\eqref{eq:4}, the interaction Hamiltonian $\Delta H$ is as follows,

\begin{equation}
    \Delta H = s\mu\left( \alpha_{y}E_{x} - \alpha_{x}E_{y} \right) + \boldsymbol{\mu} \cdot \boldsymbol{\Tilde{B}}
    \label{eq:9}
\end{equation}

\noindent According to the current configuration, in order to ensure computational simplicity and confining the electric field to two dimensions, we define a new electric field $\Tilde{\boldsymbol{E}}(\boldsymbol{r},t)=\boldsymbol{E} \times \widehat{\boldsymbol{z}}=\widehat{\boldsymbol{x}}E_{y} - \widehat{\boldsymbol{y}}E_{x}$. In this case, confined electric field vector can be written in terms of its components in the form $\left( {\Tilde{E}}_{x},{\Tilde{E}}_{y} \right) = \left( E_{y}, - E_{x} \right)$. Thus, the interaction Hamiltonian can be rearranged as in the form of,

\begin{equation}
    \Delta H = - s\mu\boldsymbol{\alpha}_{\boldsymbol{\bot}} \cdot \boldsymbol{\Tilde{E}} +\boldsymbol{\mu} \cdot \boldsymbol{\Tilde{B}}
    \label{eq:10}
\end{equation}

\noindent The current stage is crucial as it allows the total Hamiltonian to be expressed as the sum of the free and interaction terms:

\begin{equation}
    \left\lbrack H_D + \Delta H \right\rbrack \psi=i\partial_{t}\psi
    \label{eq:11}
\end{equation}

\noindent In order to eliminate the interaction term $\Delta H$, a series of preparations are required. By using the definition $\boldsymbol{p}=-i\boldsymbol{\nabla}$, we obtain:

\begin{equation}
    \left\lbrack m\beta - i \boldsymbol{\alpha}_{\boldsymbol{\bot}} \cdot \left( \boldsymbol{\nabla} - i s\mu \Tilde{\boldsymbol{E}} \right) \right\rbrack\psi=i\left\lbrack\partial_{t} + i \boldsymbol{\mu} \cdot \boldsymbol{\Tilde{B}}\right\rbrack \psi
    \label{eq:12}
\end{equation}

\noindent Here, the term $s\mu \Tilde{\boldsymbol{E}}$ is vector potential-like quantity for magnetic dipoles ($\Tilde{\boldsymbol{A}}=s\mu \Tilde{\boldsymbol{E}}$) ~\cite{Cildiroglu2018Dual, Berry1984, mignani1991}. Furthermore, in the Lorentz gauge $(-\boldsymbol{\nabla} \cdot \Tilde{\boldsymbol{A}} = \partial_t \Tilde{\phi})$, it gives rise to a scalar potential-like term $(\Tilde{\phi}= \vec{\boldsymbol{\mu}} \cdot \boldsymbol{B(t)})$. In this context, considering the motion of chargeless particles in an electromagnetic field, by incorporating the transformations of the time $(p^0)$ and spatial $(\boldsymbol{p})$ components of momentum in 2+1 dimensional spacetime:

\begin{equation}
\begin{matrix}
    p^0 \longrightarrow p^0 - \boldsymbol{\mu} \cdot \boldsymbol{B} \\
    \boldsymbol{p} \longrightarrow  \boldsymbol{p} - s \mu \Tilde{\boldsymbol{E}} 
\end{matrix}
\label{eq:13}
\end{equation}

\noindent we can achieve the covariant derivatives under this configuration:

\begin{equation}
\begin{matrix}
    i\partial_t \longrightarrow i\partial_t -  s \vec{\boldsymbol{\mu}} \cdot \boldsymbol{B} = i[\partial_t +  i s \vec{\boldsymbol{\mu}} \cdot \boldsymbol{B}] = i\Tilde{D}_t  \\
    -i\boldsymbol{\nabla}  \longrightarrow  -i\boldsymbol{\nabla} - s \mu \Tilde{\boldsymbol{E}} = -i[\boldsymbol{\nabla} - i s \mu \Tilde{\boldsymbol{E}}]=-i\Tilde{\boldsymbol{D}}
\end{matrix}
\label{eq:14}
\end{equation}

\noindent Thus, there is no magnetic field presence in the region where the particles are allowed to move, following expression is obtained for moving dipoles,

\begin{equation}
    \left\lbrack m\beta - i \boldsymbol{\alpha}_{\boldsymbol{\bot}} \cdot \Tilde{\boldsymbol{D}} \right\rbrack \psi=i\Tilde{D}_t \psi
    \label{eq:15}
\end{equation}

\noindent The form covariance of the Dirac equation requires the invariance of $\psi$:

\begin{equation}
    \left\lbrack m\beta - i \boldsymbol{\alpha}_{\boldsymbol{\bot}} \cdot \Tilde{\boldsymbol{D}}' \right\rbrack \psi' =i{\Tilde{D}}'_t \psi'
    \label{eq:16}
\end{equation}

\noindent Therefore, $\psi$ undergoes a unitary transformation as $\psi'=U\psi$, the covariant derivative defined in ~\eqref{eq:14} must transforms, 

\begin{equation}
\begin{matrix}
    D_t \psi \longrightarrow (D_t \psi )'=D'_t \psi' = U (D_t \psi) \\
    \boldsymbol{D}\psi \longrightarrow  (\boldsymbol{D}\psi)' = \boldsymbol{D}'\psi' =U(\boldsymbol{D}\psi ) 
\end{matrix}
\label{eq:17}
\end{equation}

\noindent In order to ensure unitarity, the transformation will be in the form $U=e^{i\chi(\boldsymbol{r},t)}$, where $\chi(\boldsymbol{r},t)$ is either real or Hermitian. Now, it is explicit that for ~\eqref{eq:17} to be satisfied, gauge transformations needs to be made to scalar and vector potential-like physical quantities. By considering the transformed forms of the time and spatial components of the covariant derivative, one can reach the expressions $\Tilde{\boldsymbol{A}}'=\Tilde{\boldsymbol{A}}+\boldsymbol{\nabla} \chi(\boldsymbol{r},t)$ and ${\Tilde{\phi}}'=\Tilde{\phi}-\partial_t \chi(\boldsymbol{r},t)$. After the gauge-fixing $\Tilde{\boldsymbol{A}}'=0$ and ${\Tilde{\phi}}'=0$, it is revealed that the gauge function which provides this transformation needs to be satisfy $\chi(\boldsymbol{r},t)=-s\mu\int^x \boldsymbol{E} \cdot dl + \int(\boldsymbol{\mu} \cdot \boldsymbol{B}) dt$. Here, the second term is the scalar counterpart of the AC phase, a type of scalar AB-type effect demonstrated using polarized neutrons ~\cite{Allman1992}. However, due to the nature of its characteristics, it is not of a topological structure and vanishes in the expectation value expressions which can be effectively eliminated through experimental techniques as spin-echo ~\cite{Bertlman2003}. In this sense, the non-trivial gauge function is obtained:

\begin{equation}
    \chi(\boldsymbol{r},t) = -s\mu \int^x \Tilde{\boldsymbol{E}}(\boldsymbol{r},t) \cdot d\boldsymbol{l}
    \label{eq:18}
\end{equation}

\noindent Hence, the transformed Dirac spinor is,

\begin{equation}
    \psi' = r^{is\mu\partial_t \Phi_B(t)} \psi
    \label{eq:19}
\end{equation}

\noindent This results clearly demonstrate how the variation of the magnetic field affects the interference pattern, although there is no direct interaction of the moving dipoles with the magnetic field for $r>a$. Moreover, equations ~\eqref{eq:18} and ~\eqref{eq:19} clearly indicate that the phase shares the same characteristics as other topological effects, albeit with its own distinct features. If the integral on the right-hand side of ~\eqref{eq:18} yields a linear charge distribution, the resulting effect would be an AC phase. However, incorporating the time variation of the field term indicates it would be a magnetic flux, thus described as an AB phase performed with magnetic dipoles. Here, as an illustration, let's choose the magnetic field as $\boldsymbol{B}(t)= B_0 \sin{wt} \hspace{0.1em} \boldsymbol{\hat{z}}$. Since the magnetic field is time-dependent, it induces a rotational electric field in the xy-plane: $\boldsymbol{E}(\boldsymbol{r},t)= \hspace{0.1em} \frac{ \pi a^2 w}{2r} B_0 \cos{wt} ( - \boldsymbol{\hat{x}} \sin \theta + \boldsymbol{\hat{y}} \cos\theta)
$. In this case, the transformed Dirac spinor can be obtained explicitly as  $\psi' = r^{is\mu (\pi a^2 w B_0 \cos{wt})} \psi.$

\section{Time-Dependent AB-type Effect on Electric Dipoles}

\begin{figure}
    \begin{minipage}{\linewidth}
        \centering
        \includegraphics[width=\linewidth]{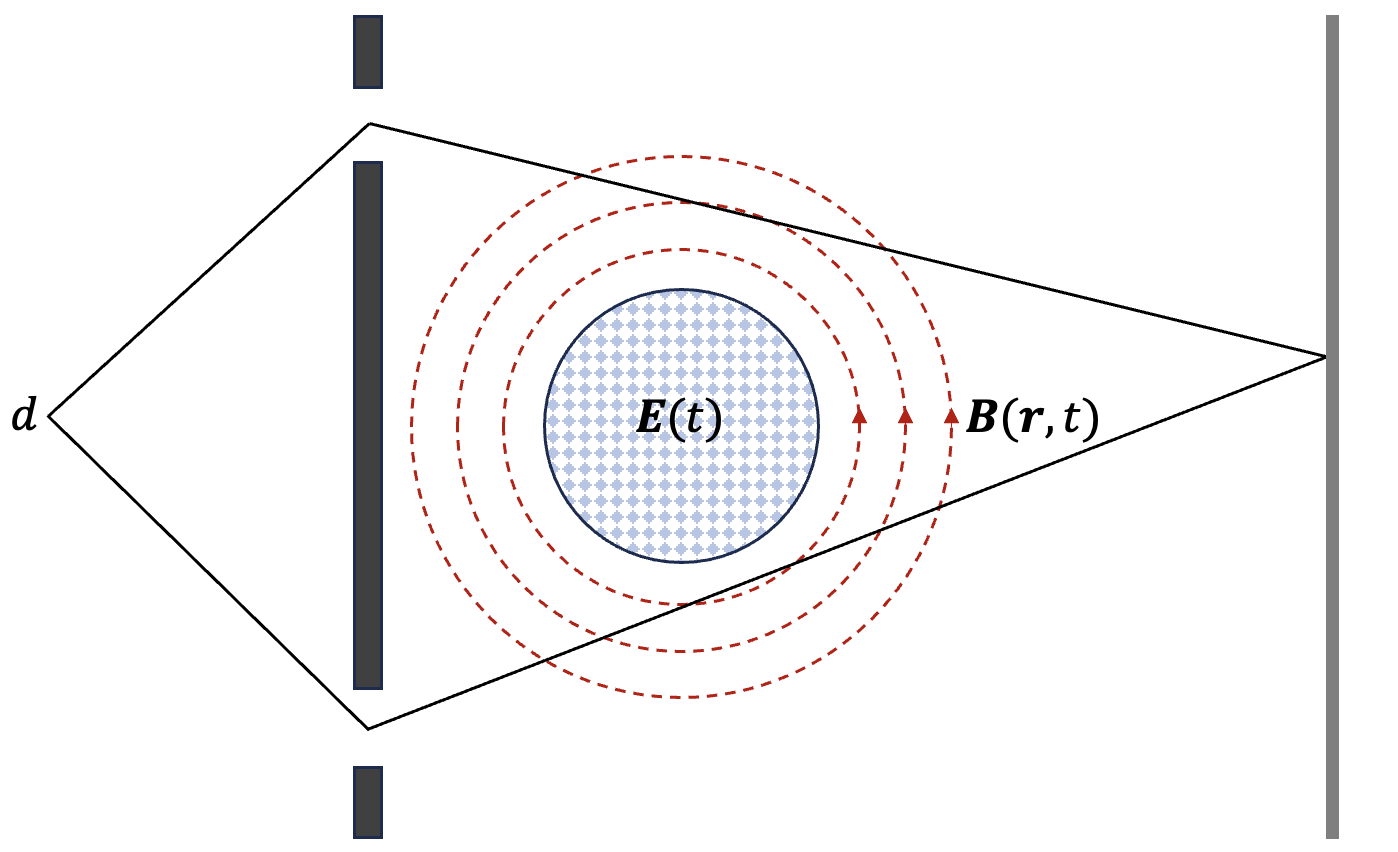}
        \caption{Schematic representation of moving electric dipoles around a time-dependent electric field source with the radius of $a$. }
        \label{fig:AB_experiment2}
    \end{minipage}
\end{figure}

In this section, as a complete dual to the discussion in the previous section, we consider chargeless particles carrying electric dipoles, moving around a time-varying electric field oriented in the z-direction within an infinitely long flux tube with radius a (See Fig.~\ref{fig:AB_experiment2}). In the region where particles are allowed to move, the presence of the time-varying electric field induces a time-dependent magnetic field ($\boldsymbol{B}(\boldsymbol{r},t)$) on the plane $(r>a)$. If the particles move in a field at velocities where first-order effects in terms of $\beta$ needs to be considered relativistic electromagnetic effects occur. Under these conditions, a vector potential-like quantity appears in the particle's wave functions as a complex phase factor, resulting in observable geometrical and topological consequences. Additionally, the induced time-dependent magnetic field generates a new electric field ($\Tilde{\boldsymbol{E}}(\boldsymbol{r},t)$) that interacts with the electric dipole, as expected for specific polarization states ~\cite{Abbott1985}. This interaction can add an extra term to the Hamiltonian, as well as a scalar phase in appropriate polarization states of the dipoles and takes the form of a scalar HMW phase contribution to the particle's wave functions ~\cite{Hashemi2019}. However, this phase is not geometrical/ topological, and can be eliminated because it cancels out in the expectation value calculations. Therefore, it is appropriate to start with the relativistic Lagrangian that governs the motion of neutral magnetic dipoles in the presence of electromagnetic fields:

\begin{equation}
    L = \overline{\psi}\left( i\gamma^{\mu}\partial_{\mu} - m - \frac{i}{2}d\sigma^{\mu\nu}\gamma^{5}F_{\mu\nu} \right)\psi
    \label{eq:20}
\end{equation}

\noindent Here, \(m\) and \(d\) are the mass and the electric dipole moment of the chargeless particles, and the \(\gamma^{5}\) is a representation formed by products of other gamma matrices defined in Clifford algebra. The equations of motion that can be derived from equation ~\eqref{eq:20} are in the form,

\begin{equation}
    \left( i\gamma^{\mu}\partial_{\mu} - m - \frac{i}{2}d\sigma^{\mu\nu}\gamma^{5}F_{\mu\nu} \right)\psi = 0.
    \label{eq:21}
\end{equation}

\noindent  The last term in parenthesis, using  ${\Tilde{F}}^{\mu\nu} = \frac{1}{2}\epsilon^{\mu\nu\alpha\beta}F_{\alpha\beta}$, can be rearranged as,

\begin{equation}
    \frac{i}{2}d\left( \sigma^{\mu\nu}\gamma_{5} \right)F_{\mu\nu} =  - \frac{d}{2}\sigma_{\alpha\beta}{\Tilde{F}}^{\alpha\beta}.
    \label{eq:22}
\end{equation}

\noindent Similar to the previous section, it is evident that the redefinition of the term $\sigma_{\alpha\beta}{\Tilde{F}}^{\alpha\beta}$ is necessary when the discussion is carried out in 2+1 dimensions. Thus, by definition $ \Sigma^{k} = \begin{pmatrix}
\sigma^{k} & 0 \\
0 & \sigma^{k} \\
\end{pmatrix}$, the term $\sigma_{\alpha\beta}{\Tilde{F}}^{\alpha\beta}$ can be explicitly written in the form,

\begin{equation}
    \sigma_{\alpha\beta}{\Tilde{F}}^{\alpha\beta} = 2i\alpha_{i}{\Tilde{F}}^{0i} + \epsilon_{ijk}\Sigma^{k}{\Tilde{F}}^{ij} = 2i\boldsymbol{\alpha}_{\boldsymbol{\bot}} \cdot \boldsymbol{B}-2\sigma_z \Tilde{E}_z
    \label{eq:23}
\end{equation}

\noindent Therefore, commencing with the initial Hamiltonian in ~\eqref{eq:4} and its representation in ~\eqref{eq:5}, when ~\eqref{eq:23} is multiplied by $\gamma^{0}$ from the left, it becomes,

\begin{equation}
    i\partial_{t}\psi = \left\lbrack m\beta + \alpha_{x}p_{x} + \alpha_{y}p_{y} + d\left( \sigma_{y}B_{x} - s\sigma_{x}B_{y} \right)-\boldsymbol{d}\cdot\boldsymbol{\Tilde{E}}\right\rbrack\psi
    \label{eq:24}
\end{equation}

\noindent The expression within the parentheses on the right-hand side represents the total Hamiltonian of the system. For ease of computation and to confine the magnetic field to the plane, we define $\Tilde{\boldsymbol{B}} = \boldsymbol{B} \times \widehat{\boldsymbol{z}}$ and $\left( \Tilde{B}{x}, \Tilde{B}{y} \right) = \left( B_{y}, - B_{x} \right)$, thereby obtaining the interaction term as follows:

\begin{equation}
    {\Delta H}=-sd\left( \alpha_{x}{\Tilde{B}}_{x} + \alpha_{y}{\Tilde{B}}_{y} \right) = - sd\boldsymbol{\alpha}_{\boldsymbol{\bot}} \cdot \Tilde{\boldsymbol{B}} -\boldsymbol{d}\cdot\boldsymbol{\Tilde{E}}
    \label{eq:25}
\end{equation}

\noindent It is clear that this term is fully dual to expression ~\eqref{eq:10}. Thus, the total Hamiltonian of the system is in the form $\left(H = H_{D}+{\Delta H}\right)$. At this stage, steps ~\eqref{eq:11}-~\eqref{eq:18} are repeated with the electric-magnetic duality transformations $\boldsymbol{\mu}\longrightarrow \boldsymbol{d}$, $\boldsymbol{E} \longrightarrow \boldsymbol{B}$, and ${\boldsymbol{B}} \longrightarrow - {\boldsymbol{E}}$, the phase is obtained as $\xi(r,t) = \text{-} sd \int^x\boldsymbol{\tilde{B}}\cdot d\boldsymbol{l} \text{-}  \int(\boldsymbol{d}\cdot \boldsymbol{\tilde{E}})dt$. Here the second term is in the form of the scalar HMW phase ~\cite{Hashemi2019}. This scalar (dynamic) phase will vanish in the expectation value expressions. Hence, the non-trivial phase and the transformed Dirac spinor are reached:

\begin{align}
        \xi(\boldsymbol{r},t)&=-sd \int^x \Tilde{\boldsymbol{B}}(\boldsymbol{r},t) \cdot d\boldsymbol{l} \nonumber \\
        \psi' &= r^{i s d \partial_t \Phi_E(t) } \psi.
    \label{eq:26}
\end{align}

\noindent The topological phase of an electric dipole under varying electric field is revealed. The electric field variation without any interaction has entered the wave function of the particles and has measurable effects on the interference pattern. The equation ~\eqref{eq:26} exhibits a complete duality with ~\eqref{eq:20}. It also shows dependence on the orientation of the spin polarizations of the particles. If the integral for the function $\xi(\boldsymbol{r},t)$ given by ~\eqref{eq:26} were to give a magnetic charge distribution, then the HMW phase would appear. However, this topological phase is completely dual to the AB-type phase introduced in the first section. In this respect, a review of the identity duality relations between the phases is left for the last section. 

\section{Discussion and Conclusion}

\begin{figure}
    \begin{minipage}{\linewidth}
        \centering
        \includegraphics[width=\linewidth]{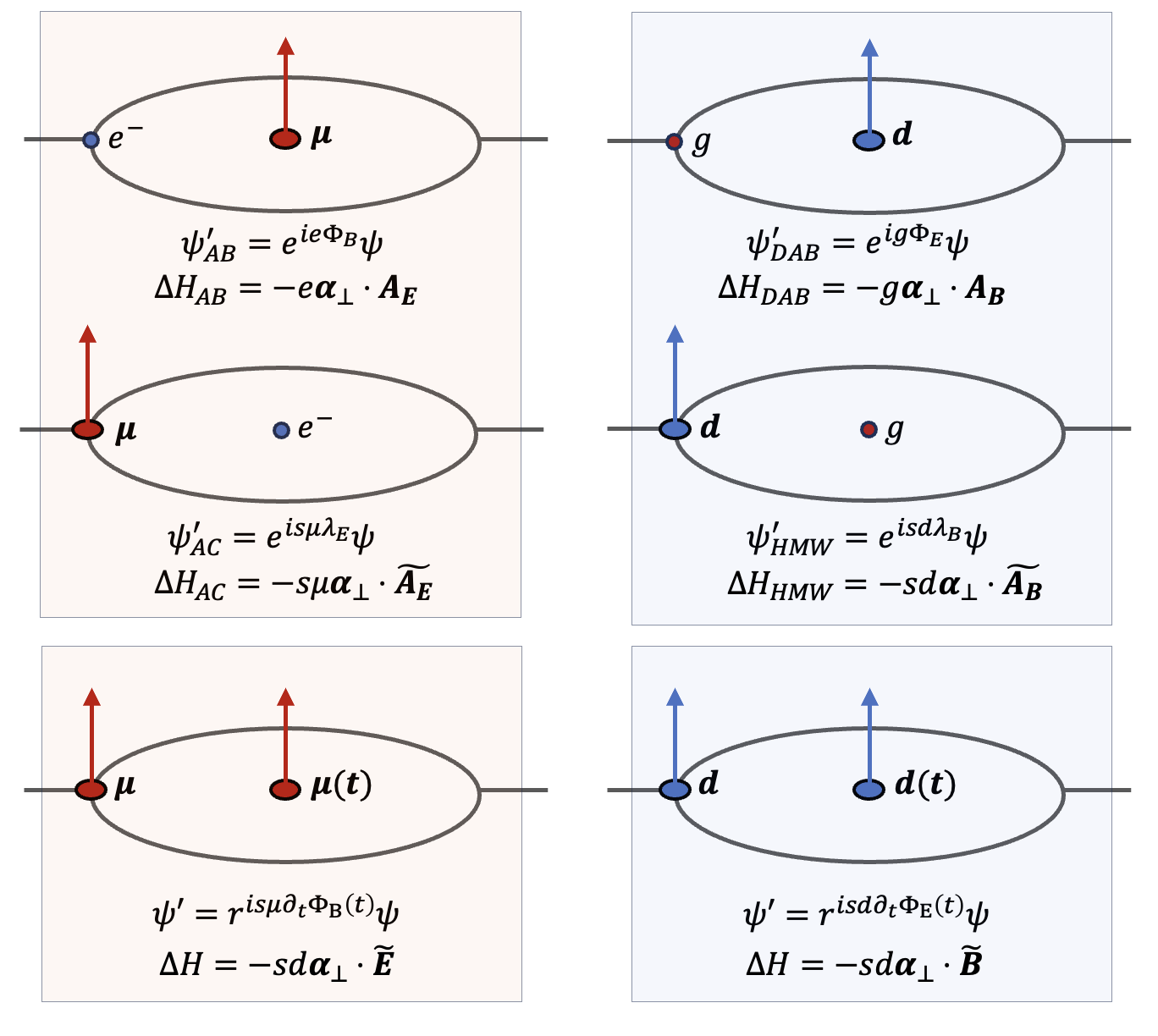}
        \caption{Schematic representations of topological phases in 2+1 dimensions. Here, the moving dipoles do not need to be polarized. This notation is utilized to reinforce the notation, and the polarization of the particles can be controlled by s. The above four phases are the four topological phases known in the literature as AB-type effects. There is a kind of identity between the AB-AC and DAB-HMW phases, provided that the moving dipoles are polarized. There is a duality relation between the left and right groups. These relations between the phases are explicit in the Hamiltonians obtained in 2+1 dimensions. Two setups given in the lower part represent schematizations of the topological phases proposed in this work. Unlike the other phases, in the condition of time dependence of the field sources, it arises as a result of the motion of the dipoles around the dipoles. Again, there is a complete duality between the two phases.}
        \label{fig:Conclusion}
    \end{minipage}
\end{figure}

AB-type effects can be addressed in 2+1 dimensions due to the symmetry of the problems. Such an investigation, conducted without approximations, illuminates the duality and identity relationships between the phases clearly. Accordingly, in two-dimensional space, the AB phase can be thought of as emerging from the motion of electrons along closed trajectories around polarized magnetic dipoles, since a solenoid can be regarded as a coherent array of magnetic dipoles (See Fig.~\ref{fig:Conclusion}). Similarly, since the electric and magnetic linear charge distributions can be considered as electric and magnetic charges in two dimensions, the AC and HMW phases emerge as a result of the motion of unpolarized magnetic dipoles around electrons and unpolarized electric dipoles around magnetic charges in closed orbits, respectively. From this perspective, one can describe an identity - rather than a duality - that emerges in the stationary reference frames of moving electrons (or polarized dipoles) between the AB and AC phases (the same relationship holds true between the DAB and HMW). However, it is clear that the AB and AC phases grouped on the left are fully dual to the DAB and HMW phases. On the other hand, AC and HMW effects depend on the polarization directions of the moving dipoles and have the same character. The AB and DAB effects do not depend on the spin orientation of the incoming electrons. All these relations can also be clearly seen from the interaction Hamiltonians describing the dynamics of the systems.

In this context, considering the scenarios proposed in the preceding sections, it becomes apparent that they correspond to topological phase structures consistent with the characteristics outlined for AB-type effects. Accordingly, the first phase, which is a combination of the AB and AC effects, is described by the motion of the moving magnetic dipoles around a variable magnetic field source. Moving particles are not required to be polarized. Their polarization can be controlled by using s, derived from representations of the Clifford algebra in 2+1 dimensions. It enters the phase factor $(\Tilde{\boldsymbol{A}}=s\mu\Tilde{\boldsymbol{E}})$ as a vector potential-like physical quantity through the electric field present region where the particles are allowed to move and the resulting relativistic effects. Thus, although there is no direct interaction, the interference pattern is measurably affected by the time-dependent variation in magnetic flux. In complete duality, a similar effect arises if electric dipoles are moved around time-dependent electric field source. Hence, the topological phase that emerges in the scenario discussed in the third section depends on the polarization states of moving particles and can be considered a combination of the HMW-DAB phases. Resultant phases align with the characteristics of AB-type effects, exhibiting complete duality between them. Furthermore, the phases are contingent upon the polarization state of the moving dipoles, and the interaction Hamiltonians take the form of AB-type effects. This opens up avenues for future studies, such as investigating the effects on entangled quantum states through the determination of instantaneous eigenvectors using matrix formalism ~\cite{Cildiroglu2019, Cildiroglu2024}.

Last, it is important to address the scalar phases that emerge as duals in the second and third sections. The phases emerging in the forms of $\chi(\boldsymbol{r},t) = -s\mu \int^x \Tilde{\boldsymbol{E}}(\boldsymbol{r},t) \cdot d\boldsymbol{l}$ and $\xi (\boldsymbol{r},t) =-sd \int^x \Tilde{\boldsymbol{B}}(\boldsymbol{r},t) \cdot d\boldsymbol{l}$ are the scalar AC ~\cite{Allman1992} and scalar HMW ~\cite{Hashemi2019} phases, respectively (For detailed information see ~\cite{Cildiroglu2018Dual}). These phases, firstly introduced to emphasize the significance of scalar potentials in physics, occur in regions without electromagnetic fields, provided they are unaffected by the classical Lorentz force. and arise from the system's dynamics under the influence of scalar potentials or similar physical quantities. Therefore, one can consider these phases as special dynamic phases that disappear in expectation value calculations and are often devoid of direct physical relevance, and can be effectively eliminated through experimental techniques such as spin-echo, thereby enabling a detailed study of the observable geometric/topological impacts of time-dependent electromagnetic field sources on dipoles.

In conclusion, the results obtained by studying the time-dependent topological effects on dipoles, in accordance with the characteristics of AB-type effects, support the complex interactions of the systems considered in the 2+1 dimension and the experimental and theoretical investigation of these interactions.

\section*{Acknowledgements}

\noindent The author would like to thank late Prof. Namık Kemal Pak for continuing to enlighten his path, Prof. Müge Boz for her invaluable contributions and precious efforts, and dear Ian Bouche for his contribution to the preliminary work. The author also extends his gratitude to the referee for their valuable guidance and unique contributions, which greatly enhanced the quality of this work. This work supported by TUBITAK BIDEB 2219 Project Number: 1059B192202412.

\bibliographystyle{elsarticle-num}
\bibliography{refs} % Burada "refs" dosyanızın adı olmalı

\end{document}